\documentclass[11pt]{article} 
\usepackage{times}
\usepackage[dvips]{epsfig}
\usepackage{a4}
\usepackage{amssymb}
\newcommand{\ceil}[1]{{\lceil{#1}\rceil}}

\begin{document}
%
\title{A Software Tool Combining Fault Masking \\
with User-Defined Recovery Strategies}
\author{Vincenzo De Florio, Geert Deconinck, Rudy Lauwereins \\
Katholieke Universiteit Leuven\\
Electrical Engineering Department, ACCA Group,\\
Kard. Mercierlaan 94, B-3001 Heverlee, Belgium.\\
\\ \\ \\ \\ \\ 
Short title of paper: Combining Fault Masking with User-Defined Recovery}
%
%
%
%
\def\today{}
\maketitle
\vfill\eject
%
\begin{abstract}
We describe the voting farm, a tool which implements a distributed software
voting mechanism for a number of parallel message passing systems.
The tool, developed in the framework of EFTOS
(Embedded Fault-Tolerant Supercomputing), can be used in stand-alone mode or
in conjunction with other EFTOS fault tolerance tools.
In the former case, we describe how the mechanism can be exploited, e.g.,
to implement restoring organs ($N\!$-modular redundancy systems with 
$N\!$-replicated voters); in the latter case, we show how it is possible 
for the user to implement in an easy and effective way a number of
different recovery strategies via a custom, high-level language. Combining
such strategies with the basic fault
masking capabilities of the voting tool makes it possible to set up complex fault-tolerant
systems such as, for instance, $N\!$-and-$M\!$-spare systems or gracefully degrading
voting farms. 
We also report about the impact that our tool can have on reliability, and
we show how,
besides structural design goals like
fault transparency, our tool achieves replication transparency, a high
degree of flexibility and ease-of-use, and good performance. 
\end{abstract}
\vfill\eject

\section{Introduction}
We herein describe the EFTOS voting farm (VF),
a class of C functions implementing a distributed software voting mechanism
developed in the framework of the ESPRIT-IV project 21012 EFTOS
(Embedded Fault-Tolerant Supercomputing)~\cite{DDLV97a,DVBD97}.

VF can be considered both as a stand-alone tool for fault masking, and
as a basic block in a more complex fault tolerance structure set up within
the EFTOS fault tolerance framework.  Accordingly, we first draw, in Sect.~\ref{standalone},
the design and the structure of the stand-alone voting farm, as a
means for orchestrating redundant resources with fault transparency as primary goal.
There, we also describe how the user can exploit the stand-alone VF tool to straightforwardly
set up systems consisting of redundant modules and based on voters, e.g.,
restoring organs (or $N\!$-modular redundancy systems with $N\!$-replicated voters).
We also report about the fault model of our tool.

In a second step (Sect.~\ref{plugged})
we concentrate on the special extra features that VF
can inherit when run as a fully EFTOS-compliant tool. To this end we 
describe the EFTOS fault tolerance layers and show how it is possible
to make use of them in order to couple the fault masking capabilities
of VF with the fault tolerance capabilities of some EFTOS tool---for instance,
adding spares to an NMR
system or allowing a voting farm to gracefully degrade in the presence
of unrecoverable faults---and we analyse in a special case the impact that
this approach can have on reliability. In the same Section we also deal with other
features that can be inherited by VF from the EFTOS framework, e.g.,
the availability of a presentation layer for hypermedia monitoring
and fault-injection.

Section~\ref{ior} deals with the impact on reliability.

Section~\ref{perf} deals with the analysis of the performance of VF
and reports also on resource consumption and some optimisations.

Section~\ref{end} concludes this work summarizing the current state of the tool. There we also
briefly describe a case study where
VF has been used in developing the software stable storage
device for the High Voltage Substations (HVS) of ENEL, the
main Italian electricity supplier.

\section{Basic structure and features of the EFTOS voting farm}\label{standalone}

\begin{figure}
\centerline{\psfig{figure=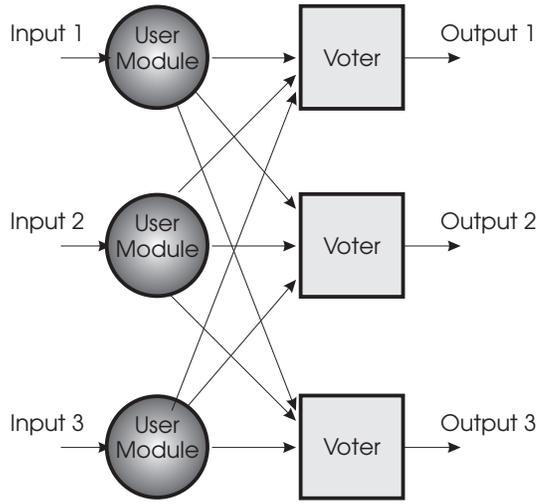,width=7.0cm}}
\caption{A restoring organ, i.e., a $N\!$-modular redundant
system with $N$ voters, when $N=3$.}
\label{ro}
\end{figure}

A well-known approach to achieve fault masking and therefore
to hide the occurrence of faults is the $N\!$-modular redundancy
technique (NMR), valid both on hardware and at software level.
To overcome the shortcoming of having one voter, whose failure
leads to the failure of the whole system even when each and every
other module is still running correctly, it is possible to use $N$
replicas of the voter and to provide $N$ copies of the inputs
to each replica, as described in~Fig.\ref{ro}.
This approach exhibits among others the following properties:
\begin{enumerate}
\item Depending on the voting technique adopted in the voter,
the occurrence of a limited number of faults in the inputs to the
voters may be masked to the subsequent modules~\cite{LoCE89};
for instance, by using majority voting, up to $\ceil{N/2}-1$ faults can
be made transparent.
\item If we consider a pipeline of such systems, then a failing voter
in one stage of the pipeline can be simply regarded as a corrupted
input for the next stage, where it will be restored.
\end{enumerate}
The resulting system is easily recognizable to be more robust
than plain NMR, as it exhibits no single-point-of-f{}ailure.
Dependability analysis confirms intuition.
Property 2. in particular explains why such systems are
also known as ``restoring organs''~\cite{John89a}.

From the point of view of software engineering, this system 
though has two major drawbacks:
\begin{itemize}
\item Each module in the NMR{} must be aware of and responsible for 
interacting with the whole set of voters;
\item The complexity of these interactions, which is a function increasing
quadratically with $N$, the cardinality of the voting farm, 
burdens each module in the NMR.
\end{itemize}

The two above observations have been recognized by us as serious
impairments to our design goals, which included
replication transparency, ease of use, and flexibility~\cite{DeDL98a}.

In order to reach the full set of our requirements, we slightly
modified the design of the system as described in Fig.~\ref{ronew}:
\begin{figure}
\centerline{\psfig{figure=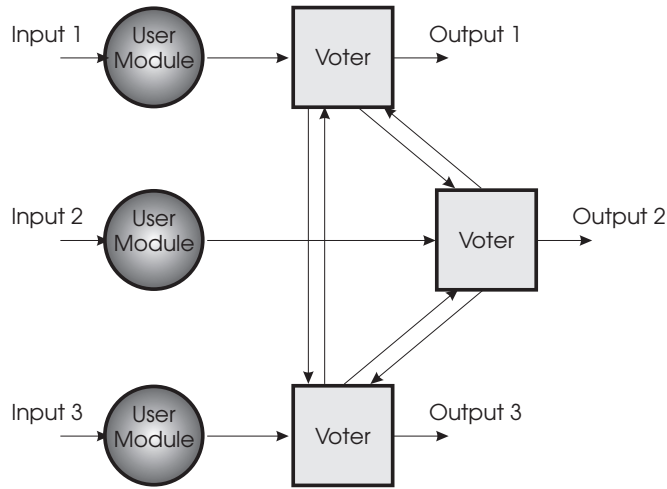,width=8.7cm}}
\caption{Structure of the EFTOS voting farm mechanism for a NMR{} system with
$N=3$ (the well-known triple modular redundancy system, or TMR).}\label{ronew}
\end{figure}
In this new picture each module only has to interact with, and be aware 
of {\em one\/} voter,
regardless the value of $N$. Moreover, the complexity of such a task
is fully shifted to the voter, i.e., transparent to the user.

\begin{figure}
\centerline{\psfig{figure=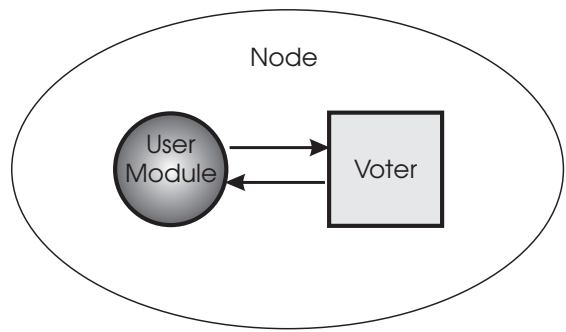,width=2.9in}}
\caption{A user module and its voter. The latter is the
only member of the farm of which the user module should be aware of:
from his/her point of view, messages will only flow between these two ends.
This has been designed so to minimize the burden of the user module
and to keep it free to continue undisturbed as much as possible.}\label{voter}
\end{figure}

The basic component of our tool is therefore the {\em voter\/} (see Fig.\ref{voter})
which we define as follows:
\begin{quote}
A voter is a local software module connected to {\em one\/}
user module and to a farm of fully interconnected fellows.
Attribute ``local'' means that both user module and voter
run on the same processing node.
\end{quote}

As a consequence of the above definition, the user module has
no other interlocutor than its voter, whose tasks are completely
transparent to the user module. It is therefore possible to model
the whole system as a simple client-server application:
on each user module the same client protocol applies (see Sect.~\ref{cs})
while the same server protocol is executed on every instance of the voter
(see Sect.~\ref{ss}).

\subsection{Client-Side of the Voting Farm: the User Module}\label{cs}
Table~\ref{example} gives an example of the client-side protocol
to be executed on each processing node of the system in which a
user module runs: a well-defined, ordered list of actions has to take place
so that the voting farm be coherently declared and defined,
described, activated, controlled, and queried:
In particular, {\em describing\/} a farm stands for
creating a static map of the allocation of its components; 
{\em activating\/} a farm substantially
means spawning the local voter (Sect.~\ref{ss} will shed more light
on this); {\em controlling\/} a farm means requesting its service
by means of control and data messages;
finally, a voting farm can also be {\em queried\/} about its state, 
the current voted value, etc.

As already mentioned, the above steps have to be carried out in the 
same way on each user module: this coherency is transparently supported
in Single-Process, Multiple-Data (SPMD) architectures. This is the
case, for instance, of Parsytec EPX
({\em Embedded Parallel eXtensions to UNIX\/}, see, e.g., ~\cite{Pars96b,Pars96a})
whose ``initial load mechanism'' transparently runs the same executable image
of the user application on each processing node of the user partition.

This protocol is available to the user as a class-like collection of
functions dealing with opaque objects referenced through pointers.
A tight resemblance with the {\sf FILE} set of functions of
the standard C language library~\cite{KeRi2} has been sought so to shorten
as much as possible the user's learning time---the
API and usage of VF closely resemble those of {\sf FILE} (see Table~\ref{cmp}).

VF has been crafted out using the CWEB system of structured
documentation~\cite{DeFl97b}, which we found to be an envaluable tool
both at design and at development time~\cite{Knuth92}.

\subsection{Server-Side of the Voting Farm: the Voter}\label{ss}
The local voter thread represents the server-side of the voting farm.
After the set up of the static description of the farm (Table~\ref{example},
Step 3) in the form of an ordered list of processing node identifiers
(positive integer numbers), the server-side of our application
is launched by the user by means of the {\sf VF\_run()} function. 
This turns the static representation of a farm 
into an ``alive'' (running) object, the voter thread.

This latter connects to its user module via inter-process communication
means (``local links'') and to the rest 
of the farm via synchronous, blocking channels (``virtual links'').
We assume the availability of a means to send and to receive messages
across these links---let us call these functions {\sf Send()} and
{\sf Receive()}. Furthermore, we assume that {\sf Send()} blocks
the caller until the communication system has fully delivered the
specified message to the specified (single) recipient, while {\sf Receive()} blocks
the caller until the communication system has fully transported
a message directed to the caller, or until a user-specified timeout has expired.

\begin{figure}
\centerline{\psfig{file=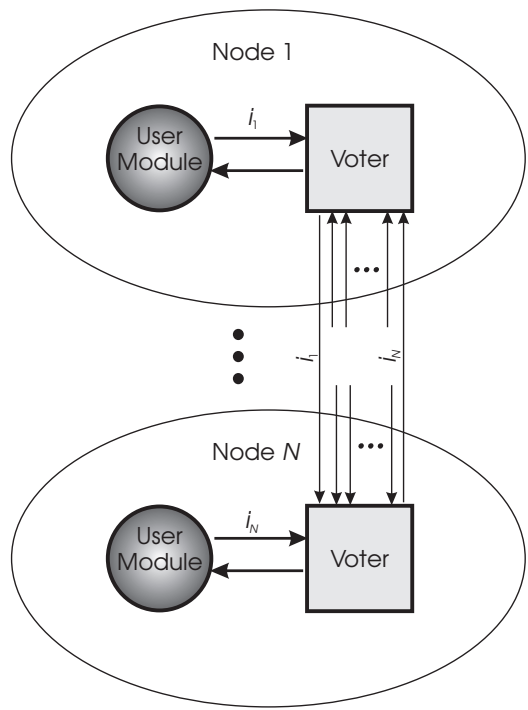,width=2.7in}}
\caption{The ``local'' input value has to be broadcast to $N-1$ fellows,
and $N-1$ ``remote'' input values have to be collected from each of the
fellows. The voting algorithm takes place as soon as a complete set of values is
available.}\label{vf}
\end{figure}

Once the connection is established, and in the absence of faults,
the voter reacts to the arrival of 
the user messages as a finite-state automaton: in particular, the arrival of
input messages triggers a number of broadcasts among the voters---as shown
in Fig.\ref{vf}---which are managed through the 
distributed algorithm described in Table~\ref{broadcast}.
Assumptions of that algorithm are fail/stop behaviour and a partially
synchronous system such that upper bounds are known for communication delays.
This last assumption is rather realistic at least in parallel environments like EPX,
which are equipped with a fast communication subsystem for their own use,
so that processors do not have to compete ``too much'' for the network.
Such subsytem also offers a reliable communication means and allows to transparently tolerate 
faults like, e.g., the break of a link, or a router's failure.

When faults occur and affect up to $M<N$ voters,
no arrival for more than $\Delta t$ time units is interpreted
as an error. As a consequence, 
variable {\sf input\_messages} is incremented as if a message
had arrived, and its faulty state is recorded. This way we can tolerate
up to $M<N$ errors at the cost of $M\Delta t$ time units.
Note that even though this algorithm tolerates up to $N-1$ faults,
the voting algorithm may be intrinsically able to cope with much less than that:
for instance, majority voting fails in the presence of faults affecting
$\ceil{N/2}$ or more voters. As another example, algorithms computing
a weighted average of the input values consider all items
whose ``faulty bit'' is set as zero-weight values, automatically
discarding them from the average. This of course may also lead to
imprecise results as the number of faults gets larger.

Besides the input value, which represents a request for voting,
the user module may send to its voter a number of other requests---some 
of these are used in Table~\ref{example}, Step 5. In particular, the user
can choose to adopt a voting algorithm among the following ones:
\begin{itemize}
\item Formalized majority voting technique,
\item Generalized median voting technique,
\item Formalized plurality voting technique,
\item Weighted averaging technique,
\item Consensus,
\end{itemize}
the first four items being the voting techniques that were generalized in~\cite{LoCE89}
to ``arbitrary $N\!$-version systems with arbitrary output types 
using a metric space framework.''
To use these algorithms, a metric function can be supplied by the
user when he/she ``opens'' the farm (Table~\ref{example}, Step 2, function {\sf objcmp()}):
this is exactly the same approach used in opaque C functions like
e.g., {\sf bsearch()} or {\sf qsort()}~\cite{KeRi2}.
A default metric function is also available.

The choice of the algorithm, as well as other control choices are managed via function
{\sf VF\_control()}, which takes as argument a voting farm pointer plus a variable number
of control argument---in Table~\ref{example}, Step 5, these arguments are
an input message, a virtual link for the output vote, an algorithm identifier, plus
an argument for that algorithm.

Other requests include the setting of some algorithmic parameters
and the removal of the voting farm (function {\sf VF\_close()}).


The voters' replies to the incoming requests are straightforward. In particular,
a {\sf VF\_DO\-NE} message is sent to the user module when a broadcast
has been performed; for the sake of avoiding deadlocks, one can only
close a farm after the {\sf VF\_DONE} message
has been sent.
Any failed attempt causes the voter to send a {\sf VF\_REFUSED}
message.
The same refusing message is sent when the user tries to
initiate a new voting session sooner than the conclusion of the previous session.

Note how function {\sf VF\_get()} (Table~\ref{example}, Step 6)
simply sets the caller in a waiting state from which it exits either on a
message arrival or on the expiration of a time-out.

\subsection{Fault Model}\label{fm}
VF can deal with the following classes of faults~\cite{Lapr95}:
\begin{itemize}
\item physical as well as human-made,
\item accidental as well as intentional,
\item development as well as operational,
\item internal and external faults,
\item permanent and temporary,
\end{itemize}

\noindent
as long as the corresponding failure domain consists only of value failures. 
Timing errors are also considered, though the delay must not be larger than some bounded value.
The tool is only capable of dealing with one fault at a time---the tool is ready to deal 
with other new faults only after having recovered from the present one.
Consistent value errors are tolerated.
Under this assumption, arbitrary in-code value errors
may occur (the adopted metric approach is not able to deal with non-code values).


\section{The voting farm as a fully EFTOS-compliant tool}\label{plugged}
The above section describes VF as a stand-alone tool. One of the consequences
of this strategy is the adoption of fault masking as a means to make
some faults transparent, e.g., via NMR{}-systems. In this Section we
now shift the attention to a more general approach towards tolerating
faults which has been defined as the object of the ESPRIT project EFTOS.
We show in particular how the conjoint use of VF and of the EFTOS framework
may yield systems that combine more than one fault tolerance technique, and
as such are more dependable.

To this end, we first introduce EFTOS and its framework, then we
describe how to make use of its extra features. Section~\ref{ior}
investigates on the impact that this version of VF can have
on reliability in a special case.

\subsection{EFTOS and its Framework}\label{ef}
The overall object of the ESPRIT-IV Project 21012 
EFTOS~\cite{DDLV97a,DVBD97} 
(Embedded Fault-Tolerant Supercomputing) has been to set up a software framework 
for integrating fault tolerance into embedded distributed high-performance
applications in a flexible, effective, and straightforward way. 
The EFTOS framework has been first implemented on a Parsytec 
CC system~\cite{Pars96a}, 
a distributed-memory MIMD supercomputer consisting of processing 
nodes based on PowerPC 604 microprocessors at 133MHz, 
dedicated high-speed links, I/O modules,
and routers. As part of the Project, this framework has been then
ported to a Microsoft Windows NT / Intel PentiumPro platform
and to a TEX / DEC Alpha platform~\cite{TEX,DEC}
so to fulfill the requirements of the EFTOS application partners.
We herein constantly refer to the version running on the CC system
and its operating system, a UNIX-dialect called EPX/nK (Parsytec's nano-kernel
version of EPX.)

The main characteristics of the CC system are the adoption of the 
thread processing model and of the message passing communication model: 
communicating threads exchange messages through a proprietary 
message passing library called EPX~\cite{Pars96b}.
The hypotheses on the target communication system we drew in Sect.~\ref{ss}
follow closely the peculiar characteristics of EPX. In particular,
EPX adopts the concepts of local and virtual links, and its {\sf Send()}
and {\sf Receive()} are compliant with what is required in Sect.~\ref{ss}.

Through the adoption of the EFTOS framework, the target embedded parallel 
application is plugged into a hierarchical, layered system 
(see Fig.~\ref{library}) whose structure and basic components are:
\begin{figure}
\centerline{\psfig{figure=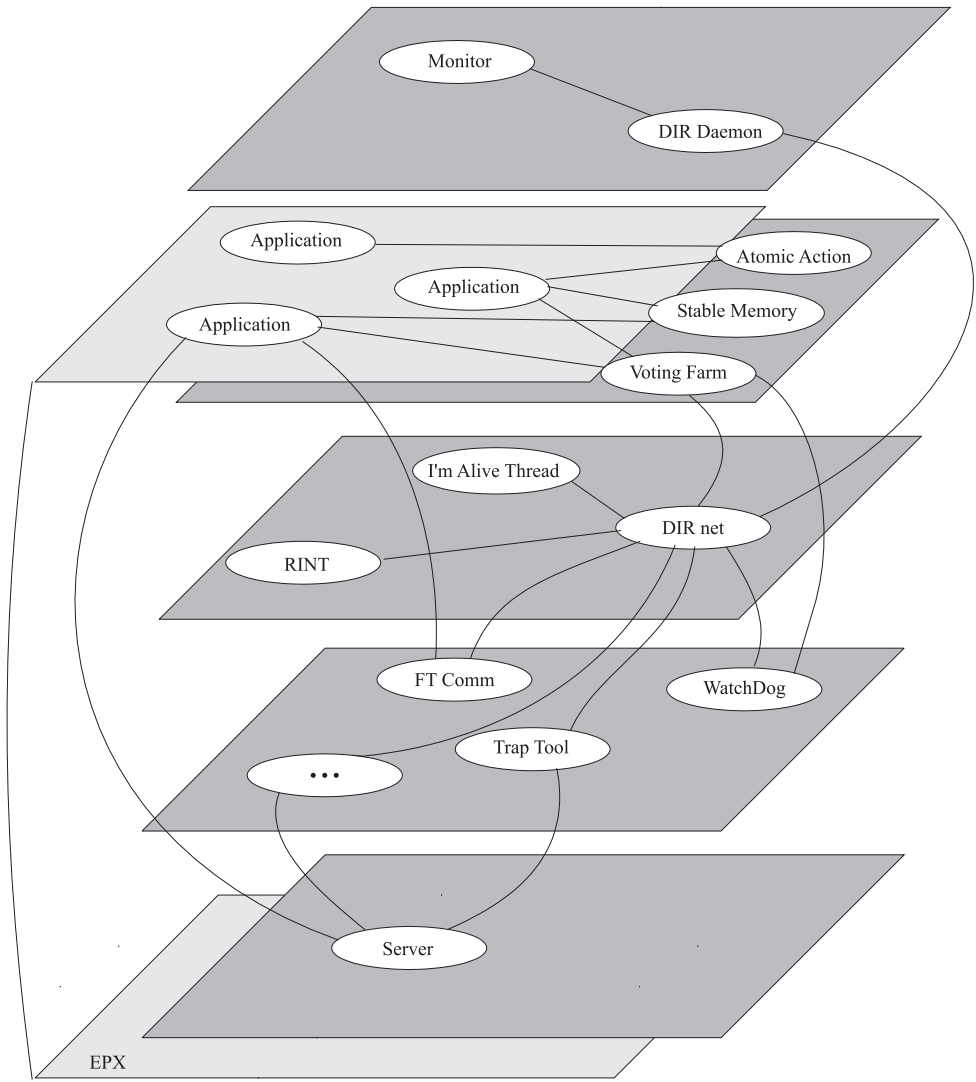,width=13.0cm}}
\caption{The structure of the EFTOS framework. Light gray has been used
for the operating system and the user application, while dark gray layers
pertain EFTOS.}
\label{library}
\end{figure}

\begin{itemize}
\item At the base level, a distributed net of ``servers'' whose
main task is mimicking possibly missing (with respect to the POSIX
standards) operating system functionalities, such as remote thread creation;

\item One level upward (detection tool layer), a set of parametrisable functions managing error
detection (we call them ``Dtools''). 
These basic components are plugged into the embedded
application to make it more dependable. EFTOS supplies a number of these
Dtools, e.g., a watchdog timer thread and a trap-handling
mechanism, plus an API for incorporating user-defined EFTOS-com\-pli\-ant tools; 

\item At the third level (control layer), a distributed application called {\em DIR net\/}
(detection, isolation, and recovery network)~\cite{TrDe97a} is available 
to coherently combine the Dtools, to ensure consistent strategies throughout the
whole system, and to play the role of a backbone handling information 
to and from the fault tolerance elements.

The DIR net can also be regarded as a fault-tolerant network of
crash-failure detectors, connected to other peripheral error detectors.
It can be used as a general toolset in which fundamental 
distributed algorithms like, e.g., those drawn in~\cite{ChTu96},
can be implemented;

\item At the fourth level (application layer), the Dtools and the components of
the DIR net are combined into dependable
mechanisms i.e., methods to guarantee fault-tolerant communication, 
stable storage devices~\cite{DBCD98},
the voting farm mechanism, etc;

\item The highest level (presentation layer) is given by
a hypermedia distributed application based on standard World-Wide Web
technology and on Tcl/Tk~\cite{Oust94}, which renders the structure and the state 
of the user application~\cite{DeDT97a}.
\end{itemize}

An important feature of the above sketched layered system is the ability to
execute user-defined recovery strategies:

\subsection{The EFTOS Recovery Language}\label{trl}
Apart from its main role of backbone of the EFTOS fault tolerance
framework, the DIR net also coordinates a virtual machine, called
the recovery interpreter (RINT for short), meant to execute
user-defined recovery strategies which are available at this level as
a set of recovery opcodes (r-codes for short)~\cite{DeFl97c}.

A special high-level language has been set up, called Recovery
Language (RL), by means of which the user can
compose custom-made recovery strategies.
Strategies are then translated into the lower level r-codes
by means of a translator. 

A RL specification is a collection of
isolation, reconfiguration, and recovery actions guarded by selective conditions ({\sf IF}'s).
Recovery actions include rebooting or shutting down a node, killing
or restarting a thread or a logical group of threads, sending
warnings to single or grouped threads, and functions to purge
error-records from the database of the DIR net.

\begin{figure}
\centerline{\psfig{file=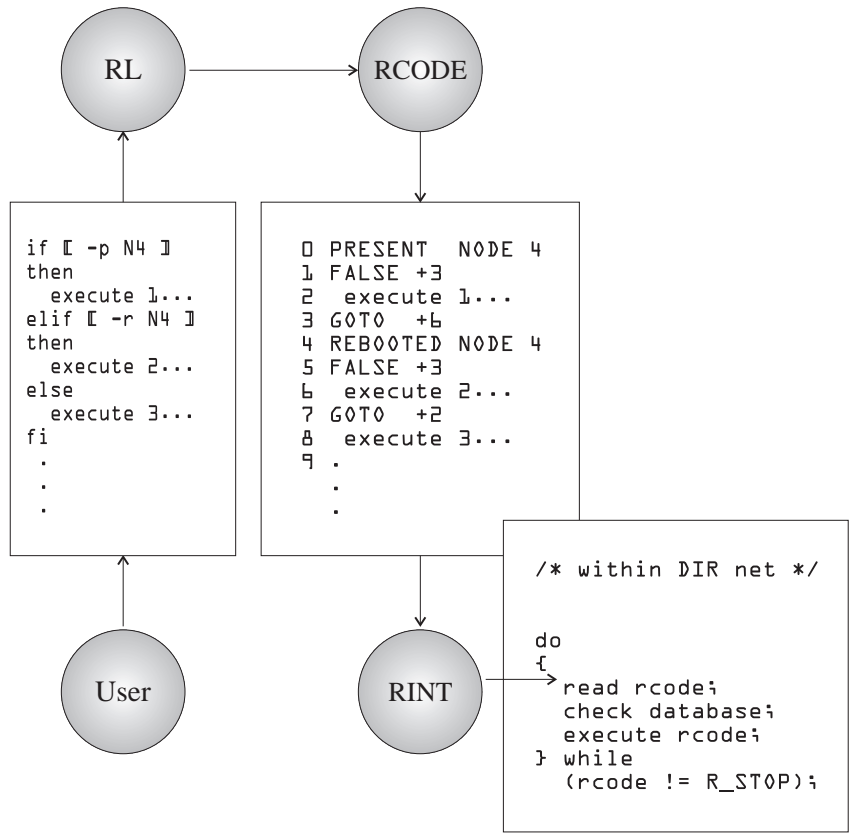,height=11cm}}
\caption{A global view of an RL program: the user supplies an RL
source code; the {\sf rl} translator turns it into binary r-codes; the r-codes
are interpreted at run time by the RINT thread within the DIR net.}
\label{scheme}
\end{figure}

The scheme works as follows: the user writes with RL a recovery strategy, 
viz. a specification of some actions to be taken to tackle each particular 
error condition as that error is detected. For each error, the DIR net awakes
the RINT thread, which interprets the r-code equivalent of the RL source,
looking for fulfilled conditions, possibly accessing the system database
maintained by the DIR net. Once a condition is met, the corresponding 
actions are executed, which are supposed to be able to tackle the error.
A default action is also available in case no {\sf IF} is evaluated as true.
The whole strategy is depicted in Fig.~\ref{scheme}.

During the lifetime of the application, this framework guards
it from a series of possible deviations from the expected
activity; this is done either by executing detection, isolation, and 
reconfiguration tasks, or by means of fault masking---this latter
being provided by the EFTOS voting farm, which we are going to describe 
in the Section to follow. 

As a last remark, the EFTOS framework appears
to the user as a library of functions written in the C programming language.

\subsection{Using the Voting Farm in Conjunction with the EFTOS Framework}
\label{rlsect}

What we described in Sect.~\ref{standalone} might be referred to as ``the Voting Farm 
in stand-alone mode'', i.e., unplugged from its originating environment,
the EFTOS framework. Here we describe further capabilities and
features that are offered to the user of our tool when run as
part of a fully EFTOS-compliant application.

When used in conjunction with the EFTOS DIR net, each voter
is transparently connected to a DIR net component and to a watch-dog timer Dtool,
connected in turn to the DIR net. This way the DIR net is informed of
errors affecting the voters. Each voter also directly informs the DIR net
of its state transitions (we call them ``phases'') by sending phase-identifiers
(pids) that can assume one of the following values:

\begin{description}
\item{{\sf VFP\_INIT},} that is, the voter is waiting for
the first input value from the user,
\item{{\sf VFP\_BROADCAST},} meaning the voter is currently
broadcasting the input value to its fellows,
\item{{\sf VFP\_VOTING},} i.e., the voter has entered the
voting function,
\item{{\sf VFP\_SUCCESS},} i.e., an output vote has been
produced and the voter is back in its waiting state;
\item{{\sf VFP\_FAILURE},} which means the voting function
was not able to produce an output vote.
\end{description}

The pid is then stored by the DIR net in its global
database~\cite{TrDe97a} and can be used during the interpretation
of the r-codes (see Sect.~\ref{trl}).

With RL the user can express strategies aiming for instance at substituting
a suspected voter with a non-faulty deputy run
elsewhere in the system (see Table~\ref{taos}), or at
a graceful degradation of the system (e.g., by killing those
voters which are in phase {\sf VFP\_FAILURE}; see Table~\ref{grade}).
Other strategies are up to the user.

Another feature that can be inherited by VF
when used as part of the EFTOS framework comes from its presentation layer:
the farm can be monitored
via the EFTOS visualization tool~\cite{DeDT97a}, a hypermedia distributed
application which remotely pilots a World-Wide Web browser so to
render the structure and the outcome of the voting sessions.

\section{Impact on Reliability}\label{ior}
Reliability can be greatly improved by this technique. For instance,
using Markov models, under the assumption of independence between faults occurrence,
it is possible to show that, let $R(t)$ be the reliability
of a single, non-replicated component, then
\begin{equation}
R^{(0)}(t) = 3R(t)^2 -2R(t)^3,\label{tmr.eq}
\end{equation}
i.e., the equation expressing the reliability of a TMR system, can be
considerably improved by adding one spare, even in the case of non-perfect
error detection coverage. 
This is the equation resulting from the Markov model in 
Fig.~\ref{markov}, expressed as a function of error recovery coverage ($C$,
defined as the probability associated with the process of identifying the failed module
out of those available and being able to switch in the spare~\cite{John89a}) and time ($t$):

\begin{figure}
\centerline{\psfig{file=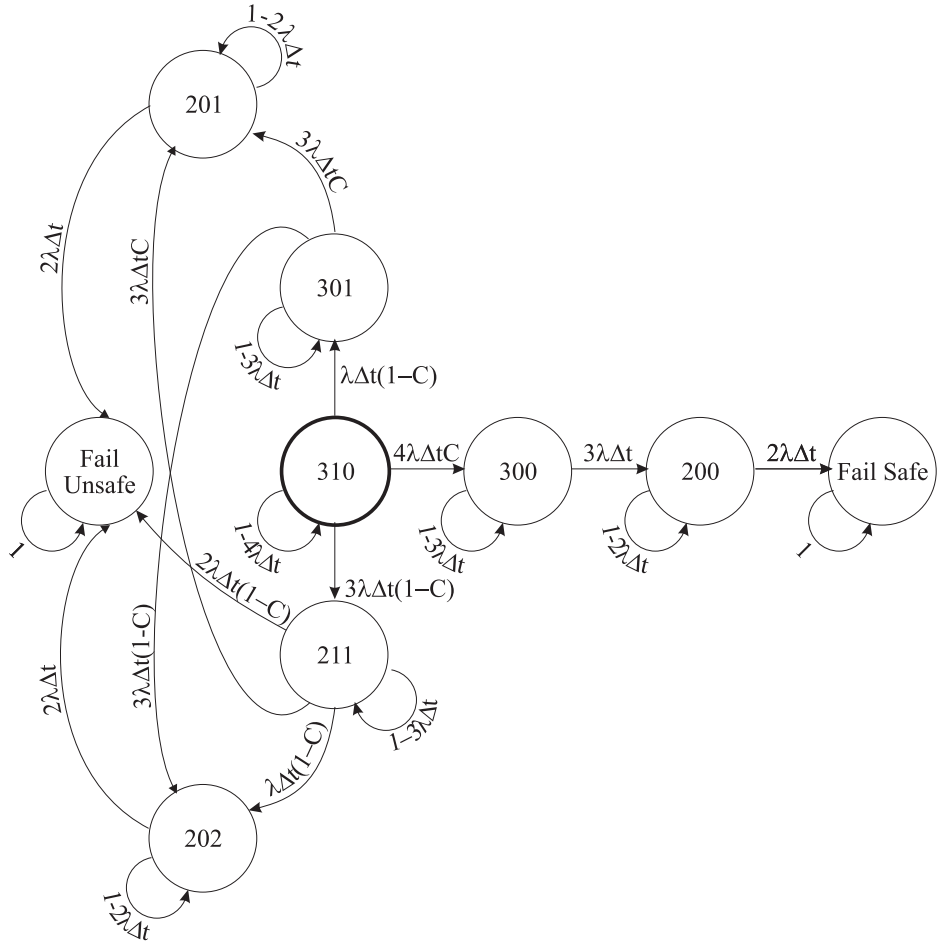,height=12.5cm}}
\caption{Markov reliability model for a TMR-and-1-spare system. $\lambda$ is the failure rate, $C$ is
the error recovery coverage factor.
`Fail safe' state is
reached when the system is no more able to correctly perform its function, though the problem has been 
safely detected and handled properly. In `Fail unsafe,' on the contrary, the system is incorrect, though
the problem has not been handled or detected. 
Every other state is labeled with three digits, $d_1d_2d_3$, such that
$d_1$ is the number of non-faulty modules in the TMR system, $d_2$ is the number of non-faulty spares
(in this case, 0 or 1), and $d_3$ is the number of undetected, faulty modules. The
initial state, 310, has been highlighted. 
This model is solved by Eq.~(\ref{ta1s.eq}).}
\label{markov}
\end{figure}

\begin{equation}
R^{(1)}(C, t) = (-3C^2 + 6C ) \times [ R(t) (1-R(t)) ]^2 + R^{(0)}(t).\label{ta1s.eq}
\end{equation}

\noindent
Appendix~\ref{a} gives some mathematical details on Eq.~(\ref{ta1s.eq}).

Adding more spares obviously implies further improving reliability. In general
we can think of a class of monotonically increasing reliability functions,
\begin{equation}
\big( R^{(M)}(C, t) \big)_{M>0},
\end{equation}
corresponding to systems adopting $N+M$ replicas.
Depending on both cost and reliability requirements,
the user can choose the most-suited values for $M$ and $N$.

\begin{figure}
\hskip-1cm
\framebox{\psfig{file=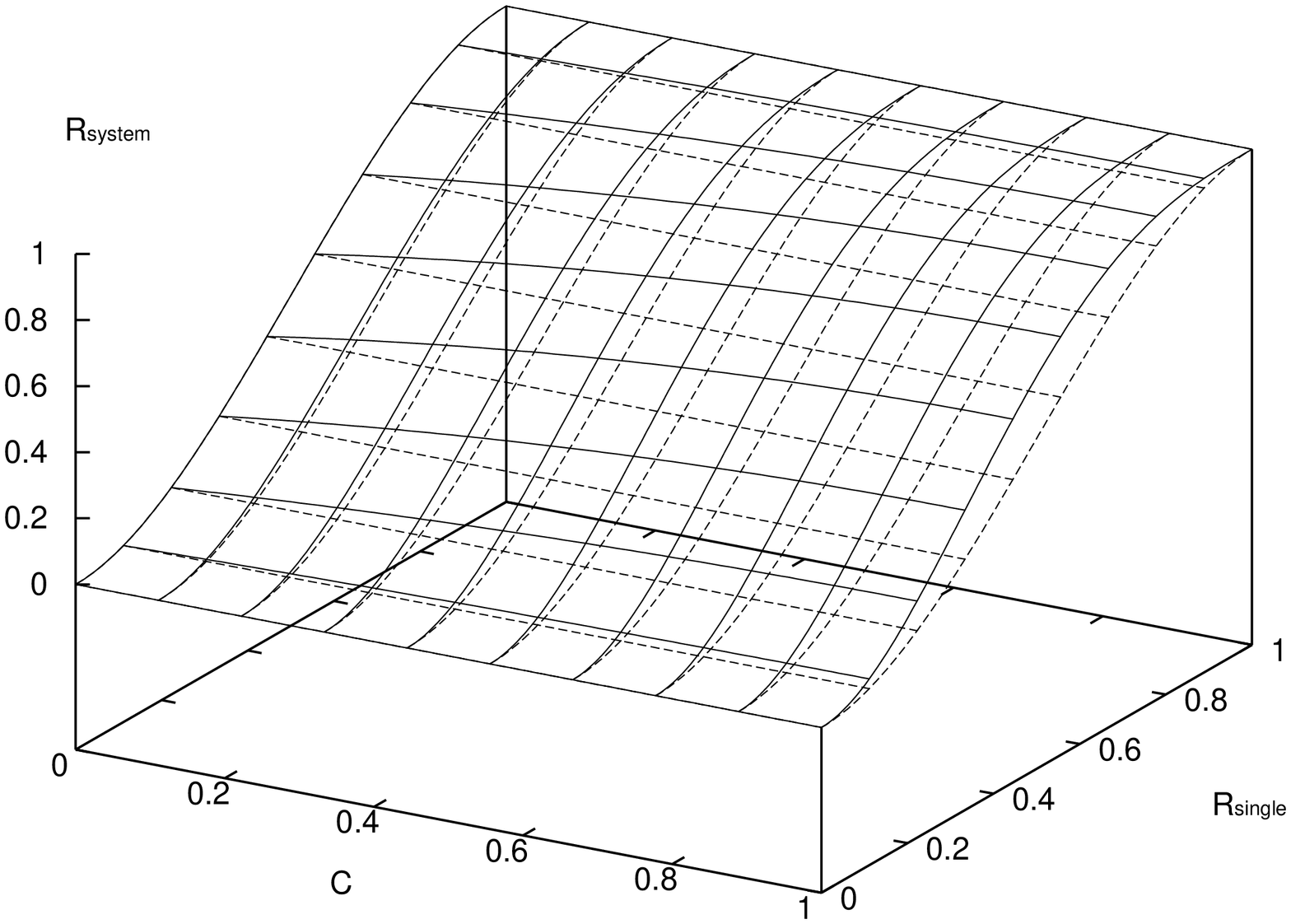,height=7.4cm,angle=-90}}%
\framebox{\psfig{file=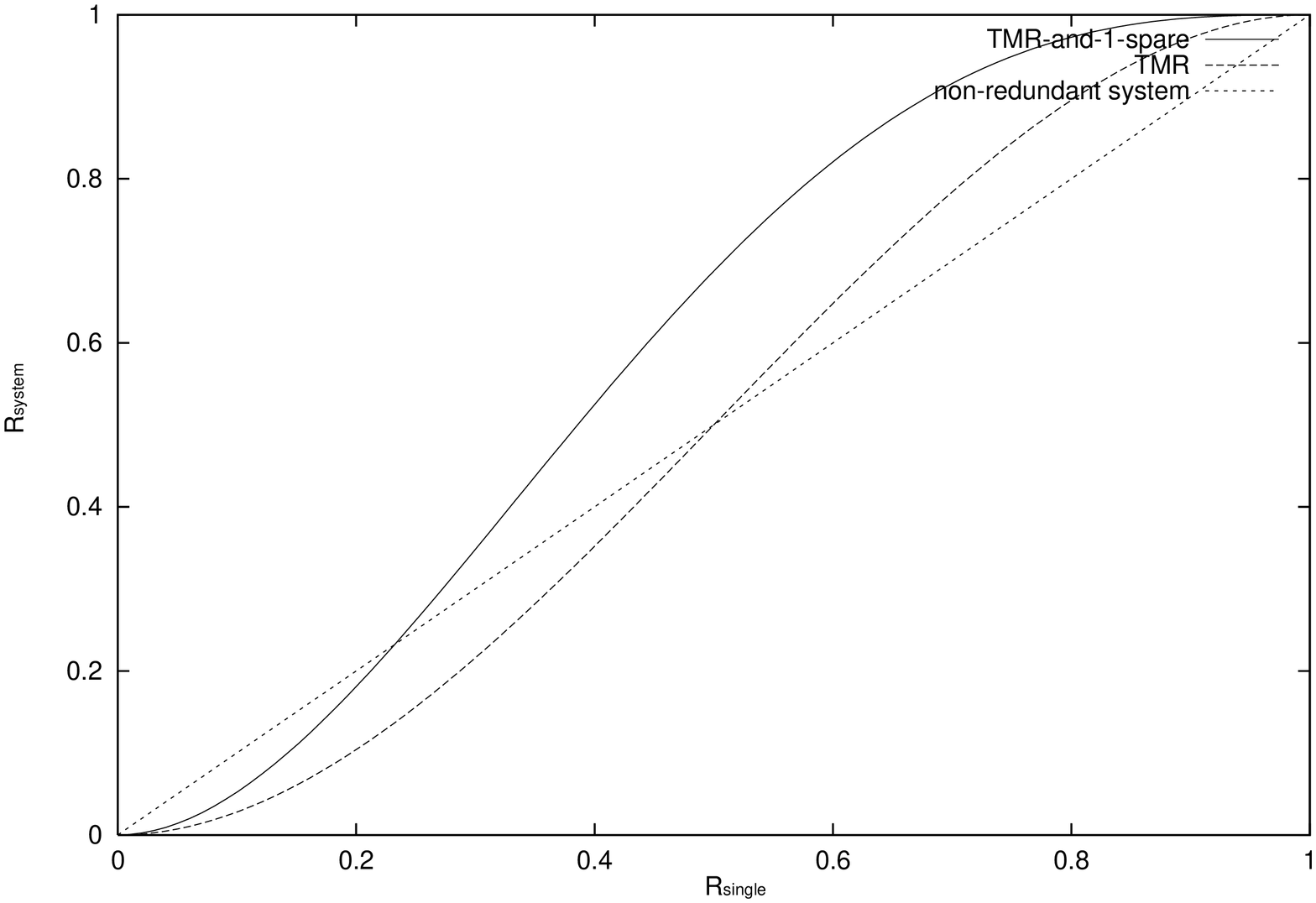,height=7.4cm,angle=-90}}
\caption{Graphs of Eq.~(\ref{tmr.eq}) and (\ref{ta1s.eq}). This latter is strictly above
the former in both pictures. The right picture considers the special case of $C=1$ (perfect
error detection coverage) and draws also
the reliability of a single, non-redundant system.}
\label{rel}
\end{figure}

Fig.~\ref{rel} compares Eq.~(\ref{tmr.eq}) and (\ref{ta1s.eq}) in the general case
(left picture) and under perfect coverage (right picture). In this latter case,
the reliability of a single, non-redundant system is also portrayed. Note furthermore how
the crosspoint between the three-and-one-spare system and the non-redundant system
is considerably lower than the crosspoint between this latter and the TMR system---$R(t)\approx 0.2324$
vs. $R(t)=0.5$.
The reliability of the system can therefore be increased
from the one of a pure NMR system to that of $N\!$-and-$M\!$-spare
systems (see Fig.~\ref{rel}).

\section{Performance of the Voting Farm}\label{perf}
This Section first focuses on estimating the voting delay via some experiments. Then it
summarises the overheads in terms of system resources (threads, memory, links\ldots)
Finally it briefly sketches a formal model used to evaluate the broadcast algorithm
in Table~\ref{broadcast} and reports on some analytical results out of it.

\subsection{Time and Resources Overheads of the Voting Farm.}
All measurements have been performed running a restoring organ consisting
of $N$ processing nodes, $N=1,\ldots,4$.
The executable file has been obtained on a Parsytec CC system
with the {\sf ancc} C compiler using the
{\sf -O} optimization flag. 
During the trials the CC system was fully dedicated to the execution
of that application. 

The application has been executed in four runs, each of which
has been repeated fifty times, increasing the number of voters from
1 to 4, one voter per node.  Wall-clock times have been collected.
Averages and standard deviations are shown in Table~\ref{run1}.

As of the overhead in resources, $N$ threads have to be spawned,
and $N$ local links are needed for the communication between each
user module and its local voter.
The network of voters calls for another 
$N\times(N-1)/2$ virtual links.

\subsection{Optimizations}
Overall performance is obviously conditioned, among other factors, by both the nature
of the broadcast algorithm as well as by the diameter
of the communication network. In particular we formally
proved~\cite{Pip} that, in a fully synchronized, fully connected (crossbar) system,
the execution time of the algorithm we adopted can vary,
depending on the permutation of the sequence of {\sf Send()}'s which constitute the
broadcast, from $O(N^2)$ (identity permutation)
to $O(N)$ (one-cycled permutation). In the case of the one-cycled permutation,
we showed that the efficiency of
the channel system is constant with respect to $N$, and approximately equal to $66.67\%$.

Due to the above results, we consider as part of the process of
porting VF to another platform
an optimization step in which a best-performing sequence
is selected. The user is also allowed
to substitute altogether the default broadcast function with another one.

\section{Conclusions}\label{end}
A flexible, easy to use, efficient mechanism for software voting 
has been described. In particular, it has been
shown how it is possible to combine fault masking with recovery techniques
when the mechanism is coupled with other tools of the EFTOS framework.
In particular, the availability of a custom, high-level language for
expressing recovery strategies allows to decouple aspects related
to fault tolerance programming from those related to a system's normal operation,
with a profitable trade-off between the need for transparent fault tolerance and
the need to orchestrate software fault tolerance in the application layer
(see e.g., ~\cite{HuKi93,Sal84}.)

The tool is currently available for a number of message passing environments, including
Parsytec EPX, Windows/NT, and TXT TEX.
A special, ``static'' version has been developed for this latter, which adopts
the mailbox paradigm as opposed to message passing via virtual links.
In this latter version, the tool has been used in a software fault tolerance 
implementation of a stable memory system (SMS) for the high-voltage substation 
controller of ENEL, the main Italian electricity supplier~\cite{DBCD98}.
This SMS is based on a combination of temporal and spatial redundancy
to tolerate both transient and permanent faults,
and uses two voting farms, one with consensus and the other with majority
voting. The tool proved to fulfill its goals and will gradually substitute
the dedicated hardware board originally implementing stable storage
at ENEL.

\paragraph{Acknowledgments.}
This project is partly supported by an 
FWO Krediet aan Navorsers, by the ESPRIT-IV
projects 21012 ``EFTOS'' and 28620 ``TIRAN,'' and by COF/96/11.
Vincenzo De Florio is on leave from the 
Tecnopolis CSATA Novus Ortus science park.
Geert Deconinck is a Postdoctoral Fellow of the 
Fund for Scientific Research - Flanders (Belgium) (FWO).
Rudy Lauwereins is a Senior Research Associate of FWO.

\vfill\eject


\vfill\eject

\appendix
\section{Appendix: Mathematical Details}\label{a}
Here we describe the basic steps leading to Eq.~\ref{ta1s.eq}.
The Markov reliability model of Fig.~\ref{markov} brings to the following set of equations:
\begin{equation}
	\left\{
	 \begin{array}{lll}
	  p_{310}(t+\Delta t) &=& p_{310}(t) (1 - 4\lambda\Delta t) \\
	  p_{300}(t+\Delta t) &=& p_{300}(t) (1 - 3\lambda\Delta t) +  p_{310}(t) 4\lambda\Delta t C \\
	  p_{200}(t+\Delta t) &=& p_{200}(t) (1 - 2\lambda\Delta t) +  p_{300}(t) 3\lambda\Delta t \\
	  p_{\hbox{\sc fs}}(t+\Delta t) &=& p_{\hbox{\sc fs}}(t) +  p_{200}(t) 2\lambda\Delta t \\
	  p_{211}(t+\Delta t) &=& p_{211}(t) (1 - 3\lambda\Delta t) +  p_{310}(t) 3\lambda\Delta t (1-C)\\
	  p_{301}(t+\Delta t) &=& p_{301}(t) (1 - 3\lambda\Delta t) +  p_{310}(t) \lambda\Delta t (1-C)\\
	  p_{201}(t+\Delta t) &=& p_{201}(t) (1 - 2\lambda\Delta t) +  p_{301}(t) 3\lambda\Delta t C +\\
	  		      & & p_{211}(t) 3\lambda\Delta t C \\
	  p_{202}(t+\Delta t) &=& p_{202}(t) (1 - 2\lambda\Delta t) +  p_{301}(t) 3\lambda\Delta t (1-C) +\\
	  		      & & p_{211}(t) \lambda\Delta t (1-C) \\
	  p_{\hbox{\sc fu}}(t+\Delta t) &=& p_{\hbox{\sc fu}}(t) +  p_{201}(t) 2\lambda\Delta t +\\
	  		      & & p_{211}(t) 2\lambda\Delta t (1-C) + p_{202}(t) 2\lambda\Delta t
	 \end{array}
	\right.\nonumber
\end{equation}

For any state $s$, let us now call $L_s = L(p_s(t))$, where $L$ is the Laplace transform.
Furthermore, assuming $(310)$ as the initial state we set $p_{310}(0)=1$ and
$\forall s \neq (310): p_s(0) = 0$. Then 
taking the limit of the above equations as $t$ goes to zero and taking the Laplace transform
we get to


\begin{equation}
	\left\{
	 \begin{array}{lll}
	  L_{310} &=& \frac{1}{s+4\lambda}\\
	  L_{300} &=& \frac{4C}{s+3\lambda} - \frac{4C}{s+4\lambda}\\
	  L_{200} &=& \frac{6C}{s+4\lambda}-\frac{12C}{s+3\lambda}+\frac{6C}{s+2\lambda}\\
	  L_{\hbox{\sc fs}} &=& \frac{C}{s}-\frac{3C}{s+4\lambda}+\frac{8C}{s+3\lambda}-\frac{6C}{s+2\lambda}\\
	  L_{211} &=& \frac{3(1-C)}{s+3\lambda} - \frac{(3(1-C)}{s+4\lambda}\\
	  L_{301} &=& \frac{1-C}{s+3\lambda} - \frac{1-C}{s+4\lambda}\\
	  L_{201} &=& 6C(1-C) (\frac{1}{s+4\lambda}-\frac{2}{s+3\lambda}+\frac{1}{s+2\lambda})\\
	  L_{202} &=& 3(1-C)^2 (\frac{1}{s+4\lambda}-\frac{2}{s+3\lambda}+\frac{1}{s+2\lambda})
	 \end{array}
	\right.\nonumber
\end{equation}

Inverting the Laplace transform we now get to
\begin{equation}
	\left\{
	 \begin{array}{lll}
	  p_{310}(t) &=& \hbox{e}^{-4\lambda t}\\
	  p_{300}(t) &=& 4C\hbox{e}^{-3\lambda t} -4C\hbox{e}^{-4\lambda t}\\
	  p_{200}(t) &=& 6C\hbox{e}^{-4\lambda t} -12C \hbox{e}^{-3\lambda t} + 6C\hbox{e}^{-2\lambda t}\\
	  p_{211}(t) &=& 3(1-C)\hbox{e}^{-3\lambda t} - 3(1-C)\hbox{e}^{-4\lambda t}\\
	  p_{301}(t) &=& (1-C)\hbox{e}^{-3\lambda t}-(1-C)\hbox{e}^{-4\lambda t}\\
	  p_{201}(t) &=& 6C(1-C) (\hbox{e}^{-4\lambda t}-2\hbox{e}^{-3\lambda t}+\hbox{e}^{-2\lambda t})\\
	  p_{202}(t) &=& 3(1-C)^2 (\hbox{e}^{-4\lambda t}-2\hbox{e}^{-3\lambda t}+\hbox{e}^{-2\lambda t})
	 \end{array}
	\right.\nonumber
\end{equation}

\noindent
(Only useful states have been computed.)

Let us call $R=\hbox{e}^{-\lambda t}$ the reliability of the basic component of the system, and
$R_{\hbox{\sc tmr}}$ the reliability of the TMR system based on the same component.
The reliability of the three and one spare system, $R^{(1)}(C,t)$, is given by the sum of the above
probabilities:

\begin{eqnarray}
R_{(1)} (C,t) &=& R^4(-3C^2+6C) + R^3(6c^2-12C-2)+R^2(-3C^2+6C+3) \nonumber\\
                        &=& (-3C^2 + 6C) (R(1-R))^2 + (3R^2 -2R^3) \nonumber\\
                        &=& (-3C^2 + 6C) (R(1-R))^2 + R_{\hbox{\sc tmr}},\nonumber
\end{eqnarray}
which proves Eq.~(\ref{ta1s.eq}).

\vfill\eject


\begin{table}[h]
\begin{small}
\begin{sf}
\vspace*{-52pt}
\begin{tabbing}
{\bf 001} \= VF\_control(vf, \=VF\_input(obj, sizeof(VFobj\_t)), \= 100000000000000000000 \= 10000000000000000000000000000 \kill\\
{\bf 1} \>  /* declaration */\\
        \> VotingFarm\_t *vf;\\
{\bf 2} \> /* definition */\\
        \> vf $\leftarrow$ VF\_open(objcmp);\\
{\bf 3} \> /* description */\\
        \> $\forall i\in\{1,\ldots,N\}$ : VF\_add(vf, node${}_i$, ident${}_i$);\\
{\bf 4} \> /* activation */\\
        \> VF\_run(vf); \\
{\bf 5} \> /* control */\\
        \> VF\_control(vf, VF\_input(obj, sizeof(VFobj\_t)), \\
	\>\>		   VF\_output(link),\\
	\>\>		   VF\_algorithm (VFA\_WEIGHTED\_AVERAGE),\\
	\>\>		   VF\_scaling\_factor(1.0) );\\
{\bf 6} \> /* query */\\
        \> do \{\} while (VF\_error==VF\_NONE \hskip10pt $\wedge$ \hskip10pt VF\_get(vf)==VF\_REFUSED);\\
{\bf 7} \> /* deactivation */\\
        \> VF\_close(vf);
\end{tabbing}
\vspace*{-15pt}
\end{sf}
\end{small}
\caption{An example of usage of the voting farm.}
\label{example}
\end{table}

\vfill\eject

\begin{table}[h]
\begin{center}
\begin{tabular}{|c|l|l|}\hline
{\bf phase}&\multicolumn{1}{|c|}{\bf{\sf FILE} class}&\multicolumn{1}{|c|}{\bf{\sf VotingFarm\_t} class}\\ \hline
declaration & {\sf FILE* f;} & {\sf VotingFarm\_t* vf;}\\ \hline
opening & {\sf f = fopen(\ldots);} & {\sf vf = VF\_open(\ldots);}\\ \hline
control & {\sf fwrite(f, \ldots);} & {\sf VF\_control(vf, \ldots);}\\ \hline
closings & {\sf fclose(f);} & {\sf VF\_close(vf);}\\ \hline
\end{tabular}
\end{center}
\caption{The C language standard class for managing file is compared with
the VF class. The tight resemblance has been sought in order to shorten
as much as possible the user's learning time.}
\label{cmp}
\end{table}

\vfill\eject

\begin{table}[h]
\begin{small}
\begin{sf}
\hrulefill
\vspace*{-12pt}
\begin{tabbing}
{\bf 11} \= 100 \= 10000000 \= 1000000000 \= 1000000000 \= 1000000 \= 1000000\kill\\
{\bf 1} \> /* each voter gets a unique voter\_id $\in\{1,\ldots,N\}$ */ \\
        \> voter\_id $\leftarrow$ who-am-i();\\
{\bf 2} \> /* all messages are first supposed to be valid */\\
        \> $\forall i$ : valid${}_i \leftarrow$ TRUE; \\
{\bf 3} \> /* keep track of the number of received input messages */\\
        \> $i\leftarrow$ input\_messages $\leftarrow$ 0;\\
{\bf 4} \> do \{ \\
{\bf 5} \>\> /* wait for an incoming message or a timeout */\\
        \> \> Wait\_Msg\_With\_Timeout($\Delta t$);\\
{\bf 6} \>\>/* $u$ points to the user module's input */\\
        \>\> if ( Sender() $\equiv$ USER ) $u \leftarrow i$; \\
{\bf 7} \>\>/* read it */\\
        \>\> if ( $\neg$ Timeout ) msg${}_i \leftarrow$ Receive(); \\
{\bf 8} \>\>/* or invalidate its entry */\\
        \>\> else valid${}_i \leftarrow$  FALSE; \\
{\bf 9} \>\>/* count it */\\
        \>\> $i\leftarrow$ input\_messages $\leftarrow$ input\_messages + 1;\\
{\bf 10} \> \> if (voter\_id $\equiv$ input\_messages) Broadcast(msg${}_u$);\\
{\bf 11}\> \} while (input\_messages $\neq$ N);
\end{tabbing}
\vspace*{-5pt}
\hrulefill
\end{sf}
\end{small}
\caption{The distributed algorithm needed to regulate the right to
broadcast among the $N$ voters. Each voter waits for a message
for a time which is at most $\Delta t$, then it assumes a fault affected
either a user module or its voter. Function {\sf Broadcast()} 
sends its argument to all voters whose id is different from
{\sf voter\_id}. It is managed via a special sending thread so to
circumvent the case of a possibly deadlock-prone {\sf Send()}.}
\label{broadcast}
\end{table}

\vfill\eject

\begin{table}[h]
\begin{sf}
\hrulefill
\begin{tabbing}
{\bf 000}\=THEN\=THEN\kill
\> INCLUDE "vf\_phases.h"\\
\> IF [ -FAULTY THREAD1  \\
\> \hskip15pt OR  -PHASE THREAD1  ==  $\{$VFP\_FAILURE$\}$ ]\\
\> THEN\\
\>\>   KILL THREAD1\\
\>\>   START THREAD4  AND \\
\>\>  \hskip15pt WARN  THREAD2, THREAD3\\
\> FI
\end{tabbing}
\hrulefill
\end{sf}
\caption{A recovery rule coded in RL.
We suppose a voting farm consisting of three threads, identified by
integers 1--3. If the first thread is detected as faulty or its state is
{\sf VFP\_FAILURE}, then that thread is killed, a new thread is started
and the fellows of the faulty one are alerted so that they restore
a non-faulty farm. Three of such rules may be used to
set up, e.g., a three-and-one-spare system. (Note the {\sf INCLUDE} statement,
which is used to import C-style definitions into RL. Note also the curly
brackets operator, which de-references such definitions, as in
{\sf $\{$VFP\_FAILURE$\}$}.)}
\label{taos}
\end{table}

\vfill\eject

\begin{table}[h]
\begin{sf}
\hrulefill
\begin{tabbing}
{\bf 000}\=THEN\=THEN\kill
\> INCLUDE "vf\_phases.h"\\
\> IF [ -FAULTY GROUP1  \\
\> \hskip15pt OR  -PHASE GROUP1  ==  $\{$VFP\_FAILURE$\}$ ]\\
\> THEN\\
\>\>   KILL THREAD@  AND WARN  THREAD$\sim$
\\
\> FI
\end{tabbing}
\hrulefill
\end{sf}
\caption{Another recovery rule coded in RL.
We suppose a voting farm consisting of three threads, collectively
identified as ``group 1.'' The {\sf IF} statement checks whether
any element of the group has been detected as faulty or is
currently in {\sf VFP\_FAILURE} state. If so, in the first
action, those that fulfill
the condition (identified in RL as {\sf THREAD@}) are killed, while
those that do not fulfill the condition (in RL, {\sf THREAD$\sim$})
are warned.  This allows a graceful degradation of the voting farm.}
\label{grade}
\end{table}

\vfill\eject

\begin{table}[h]
\centerline{\begin{tabular}{|r|c|c|}
\hline
number of nodes	& \ \ average\ \ \ & standard deviation \\ \hline
\hbox to 25pt{1}	& 0.000615	& 0.000006 \\ 
\hbox to 25pt{2}	& 0.001684	& 0.000022 \\ 
\hbox to 25pt{3}	& 0.002224	& 0.000035 \\ 
\hbox to 25pt{4}	& 0.003502	& 0.000144 \\ 
\hline
\end{tabular}
}
\caption{Time overhead of the voting farm for one to four node systems (one voter
per node). The unit is seconds.}
\label{run1}
\end{table}

\end{document}